\documentclass[aps,twocolumn,pra,showpacs,floatfix]{revtex4}
\usepackage{epsfig}
\usepackage{graphicx}
\usepackage{dcolumn}

\begin{document}

\title{Current trends in searches for new physics using measurements of 
 parity violation and electric dipole moments in atoms and molecules}

\author{V. A. Dzuba}
\email[E-mail address: ]{v.dzuba@unsw.edu.au}
\affiliation{School of Physics, University of New South Wales, Sydney 2052,
Australia}
\author{V. V. Flambaum}
\email[E-mail address: ]{v.flambaum@unsw.edu.au}
\affiliation{School of Physics, University of New South Wales, Sydney 2052,
Australia \\ and\\ European Centre for Theoretical Studies in
Nuclear Physics (ECT), Strada della Tabarelle 286, I-38123, Villazzano
(Trento), Italy}

\begin{abstract}
We review current status of the study of parity and time invariance
phenomena in atoms, nuclei and molecules. We focus on three most
promising areas of research: (i) parity non-conservation in a chain of
isotopes, (ii) search for nuclear anapole moments, and (iii) search
for permanent electric dipole moments (EDM) of atoms and molecules
which are caused by either, electron EDM or nuclear $T,P$-odd moments
such as nuclear EDM and nuclear Schiff moment. 
\end{abstract}


\maketitle

\section{Introduction}

The study of the parity and time invariance violation in atoms,
molecules and nuclei is a low-energy, relatively inexpensive
alternative to high-energy search for new physics beyond the standard
model (see, e.g. a review~\cite{Ginges}). Until very recently the
accurate measurements of the parity non-conservation (PNC) in atoms
was one of the most promising ways of exploring this path. It 
culminated in very precise measurements of the PNC in
cesium~\cite{Wood}. There were even indications that these
measurements show some disagreements with the standard model and might
indeed lead to new physics~\cite{BW}. It took considerable effort
of several groups of theorists to improve the interpretation of
the measurements and resolve the disagreement in
favor of the standard model. The disagreement were removed when
Breit~\cite{Breit} and quantum electrodynamic corrections~\cite{QED}
were included and the accuracy of the treatment of atomic correlations
were improved~\cite{DFG02,PBD09}. 

In is unlikely that any new measurements of the PNC in atoms can
compete with the cesium experiment in accuracy of the measurements and
interpretation (the heavy  alkaline atoms like Fr and Ra$^+$,  where
the PNC effect is 20 times larger than in Cs, may be  exceptions).
Therefore, the main interest in the area has shifted
mostly to three other important subjects: (i) the PNC measurements for a
chain of isotopes; (ii) the measurements of nuclear anapole moments; and
(iii) the measurements of the P,T-odd permanent electric dipole moments
of atoms and molecules. Below we will briefly review each of these
subjects. 

\section{Chain of  isotopes}

The values measured in atomic PNC-experiments can be presented in a
form
\begin{equation}
 E_{PNC} = k_{PNC} Q_W,
\label{eq:Epnc}
\end{equation}
where $k_{PNC}$ is an electron structure factor which comes from atomic
calculations, and $Q_W$ is a weak nuclear charge. Very sophisticated
calculations are needed for accurate interpretation of the
measurements as has been discussed in introduction for the case of
Cs. An alternative approach was suggested in
Ref.~\cite{DFK86}. If the same PNC effect is measured for at least two
different isotopes of the same atom than the ratio
\begin{equation}
\mathcal{R} =\frac{E'_{PNC}}{E_{PNC}} = \frac{Q'_W}{Q_W}
\label{eq:R}
\end{equation}
of the PNC signals for the two isotopes does not depend on electron
structure factor. It was pointed out however in Ref.~\cite{Fortson}
that possible constrains on the new physic coming from isotope ratio
measurements are sensitive to the uncertainties in the neutron
distribution which are sufficiently large to be a strong limitation
factor on the value of such measurements. The problem was addressed in
Ref.~\cite{DP02} and more recently in Ref.~\cite{BDF09}. The authors
of Ref.~\cite{DP02} argue that experimental data on neutron
distribution, such as, e.g. the data from the experiments with
antiprotonic atoms\cite{Trzcinska}, can be used to reduce the
uncertainty. In more general approach of Ref.~\cite{BDF09} nuclear
calculations are used to demonstrate that the neutron distributions
are correlated for different isotopes which leads to cancelation of
the relevant uncertainties in the ratio (\ref{eq:R}).

The parameter $\mathcal{F}$ of the sensitivity of the ratio
(\ref{eq:R}) to new physics can be presented in the form
\begin{equation}
\mathcal{F}= \frac{h_p}{h_0} = \left( \frac{\mathcal{R}}{\mathcal{R}_0}
  -1 \right) \frac{NN'}{Z\Delta N},
\label{eq:F}
\end{equation}
where $h_p$ is the new physics coupling to protons ($\Delta Q_{new} =
Zh_p+Nh_n$), $h_0$ comes from the SM, $\mathcal{R}_0$ is the ratio
(\ref{eq:R}) assuming that each isotope has the same proton and
neutron distribution (no neutron skin),
$N$ and $N'$ are the numbers of neutrons in two isotopes, $Z$ is the
number of protons and $\Delta N=N'-N$.
The constrains on new physics parameter $h_p$ are affected by the
experimental error $\delta \mathcal{R}_{exp}$ and uncertainties in
$\mathcal{R}_0$ due to unsufficient knowledge of nuclear
distributions. The later, as is argued in Ref.~\cite{BDF09}, are
correlated and strongly cancel each other. Estimations of
Ref.~\cite{BDF09} show that corresponding contribution to
$\delta\mathcal{F}$ is in the range $10^{-3} \div 10^{-2}$ which is
about an order of magnitude smaller than the uncorrelated ones. 
In the end, the isotope-chain measurements are more sensitive to new
physics than current parity-violating electron scattering
measurements~\cite{PVES} (by a factor of 10 for such atoms as Cs, Ba
and Dy). 

Experiments on isotope chains are in progress at Berkeley for Dy and
Yb atoms \cite{DyPNC,YbPNC}, at TRIUMF for Fr atoms~\cite{FrPNC},
at Los Alamos for Yb$^+$ ions~\cite{Yb+PNC},  and at Groningen (KVI)
for Ra$^+$ ions~\cite{KVI}.

\section{Anapole moment}

The notion of the anapole moment was introduced by
B. Ya. Zeldovich~\cite{zeldovich1957}. 
Nuclear anapole moment (AM) is the magnetic $P$ and $C$-odd, $T$-even
nuclear moment caused by the $P$-odd 
weak nuclear forces. 
Interaction of electrons with AM magnetic field (which may be called the
PNC hyperfine interaction) dominates the
nuclear-spin-dependent contribution to the atomic or molecular PNC effect.

First calculations of nuclear AM and proposals for
experimental measurements were presented in
Ref.~\cite{FKh80,FKhS84,SF78,FKh85}. Corrections to the AM interaction
with electrons due to
finite nuclear size were considered in Ref.~\cite{FH93}.
The authors of \cite{SF78,FKh85} (see also \cite{labzovsky1978}) note
in particular 
that the effect of AM is strongly enhanced in diatomic
molecules due to mixing of the close rotational
states of opposite parity including mixing of $\Lambda$ or $\Omega$ doublets. 
The PNC effects produced by the weak charge are not enhanced. Therefore
the AM effect dominates PNC in molecules.
This greatly simplifies the detection of AM in diatomic molecules
compared to atoms. In atoms the AM effect is 50 times
smaller than the weak charge effect; AM effect is separated as a small
difference of the PNC effects in different hyperfine transitions.
A review of the parity and time invariance violation in diatomic
molecules (including the AM effect) can be found in Ref.~\cite{KL95}. 

The idea
of the AM contribution enhancement may be explained as follows. 
After the averaging over electron wave function the effective operator
acting on the angular variables may contain three vectors:
the direction of molecular axis ${\bf N}$, the electron angular
momentum ${\bf J}$
and nuclear spin ${\bf I}$. Scalar products ${\bf NI}$ and 
${\bf NJ}$ are $T$-odd and  $P$-odd. Therefore, they
are produced by the $T,P$-odd interactions discussed in the next section.
$P$-odd, $T$-even operator $V_P$ must be proportional to ${\bf N [J \times
  I]}$. It contains nuclear spin ${\bf I}$, therefore, the 
weak charge does not
contribute. The nuclear AM is directed along the nuclear
spin ${\bf I}$, therefore, it contributes to $V_P$. The matrix elements of 
${\bf N}$ between molecular rotational states are well-known, they produce
rotational electric dipole transitions in polar molecules. Therefore,
$V_P$ (induced by the magnetic interaction of the nuclear AM with
molecular electrons) mixes close  rotational-hyperfine states of opposite
parity.  
The interval between these levels is five orders of magnitude smaller
than the interval between the opposite parity levels in atoms
(by the factor $m_e/M$ where $m_e$ and $M$ are the electron and
reduced molecular masses), therefore PNC effects are five orders
of magnitude larger. 
Further enhancement may be achieved by a reduction of the intervals by an
external magnetic field~\cite{FKh85}.

The effect is further enhanced for heavy
molecules. It grows with nuclear charge as $Z^2A^{2/3}R(Z\alpha)$,
where $R(Z\alpha)$ is the relativistic factor which grows from $R=1$ at
low $Z$ to $R \sim 10$ for $Z > 80$. Good candidates for the
measurements include the molecules and molecular ions with
$\Sigma_{1/2}$ or $\Pi_{1/2}$ electronic ground
state~\cite{SF78,FKh85}, for example,  YbF, BaF, HgF, PbF, LaO,
LuO, LaS, LuS, BiO, BiS, YbO+, PbO+, BaO+, HgO+, etc.
Molecular experiments are currently
in progress at Yale~\cite{DeMille} and Groningen KVI~\cite{Jangman}.
An interesting idea of studying AM contribution to the NMR spectra of
chiral molecules were discussed in Ref.~\cite{Nahrwold}.

So far the only nuclear AM which has been measured is the AM of the
$^{133}$Cs nucleus. It is done by comparing PNC
amplitudes between different hyperfine structure sublevels in the same
PNC experiment which is discussed in the
introduction~\cite{Wood}. Interpretation of the
measurements~\cite{Murray} indicates 
some problems. 
For example, the value of Cs AM is inconsistent with the limit on the AM
of Tl~\cite{Tlanapole}.

To resolve the inconsistencies and obtain valuable information about
P-odd nuclear forces it would be extremely important to measure
anapole moments for other nuclei. In particular, it is important to
measure AM for a nucleus with an unpaired neutron (Cs and
Tl have unpaired protons). AM of such nucleus depend on different
combination of the weak interaction constants providing important
cross-check. Good candidates for such measurements include odd isotopes of
Ra, Dy, Pb, Ba, La, Lu and Yb. The Ra atom has an extra advantage
because of strong 
enhancement of the spin-dependent PNC effect in the $^1$S$_0$ -
$^3$D$_2$ transition due to proximity of the
opposite-parity state $^3$P$^o_1$ ($\Delta E=5$ cm$^{-1}$)~\cite{Ra}.

Experimental work is in progress for Rb and Fr at
TRIUMF~\cite{RbAM,FrPNC}, 
and for Dy and Yb at Berkeley~\cite{DyPNC,YbPNC}.  

\section{Electric dipole moment}

Permanent electric dipole moment (EDM) of a neutron, atom or molecule
would violate both $P$ and $T$ invariance. Under conditions of the
$CPT$-theorem this would also mean a $CP$-violation. The
Kobayashi-Maskawa mechanism of the SM leads to extremely small values
of the EDMs of the particles. It is also too weak to explain the
matter-antimatter asymmetry of the Universe. On the other hand, most
of the popular extensions to the SM predict much larger EDMs which
are within experimental reach.
The EDM of an atom or a molecule is mostly due to
either electron EDM and T,P-odd electron-nucleon interactions in
paramagnetic systems (with non-zero total momentum $J$) or to the
$T,P$-odd nuclear forces in diamagnetic systems 
($J=0$; nuclear-spin-dependent e-N interaction contributes here too).
The existence of $T,P$-odd nuclear forces leads to the $T,P$-odd
nuclear moments in the expansion of 
the nuclear potential over powers of distance $R$ from the center of
the nucleus. The lowest-order term in the expansion, the nuclear EDM,
is unobservable in neutral atoms due to total screening of the
external electric field by atomic electrons. It might be possible
however to observe the nuclear EDM in ions (see below). The first
non-vanishing term which survives the screening in neutral systems is
the so called Schiff moment. Below we discuss the effects of nuclear
and electron EDM and the Schiff moment.

\subsection{Nuclear EDM}

It was widely believed that one needs neutral particles (e.g.,
neutron, neutral atom or molecule) to study EDMs. This is because the
EDM is expected to be very small and it would be very hard to see the
effect of its interaction with external electric field on the
background of the much stronger interaction with the electric
charge. On the other hand, the EDM of neutral systems is very much
suppressed by the effect of screening of the external electric field by
electrons (Schiff theorem). 
The Schiff theorem may be violated by the relativistic effect (which
dominates in the case of the electron EDM), hyperfine interaction and finite
size effect. For example, the lowest-order $T,P$-odd nuclear moment, the
nuclear EDM is practically unobservable in the neutral systems (except for a
small contribution due to the hyperfine interaction). First observable $T,P$%
-odd nuclear moment, the Schiff moment, is non-zero due to finite nuclear
size.

It is important therefore to explore the possibility of studying EDMs of
charged particles (e.g. muons or atomic ions). There are realistic
suggestions of this kind in Refs.~\cite{Farley,Orlov,Baryshevsky}
based on the use of ion storage rings.

The external electric field is not totally screened on the nucleus of
an ion. Its value is
\begin{equation}
  E_N = \frac{Z_i}{Z} E_0,
\label{eq:EN}
\end{equation}
where $E_0$ is external electric field, $E_N$ is electric field at the
nucleus, $Ze$ is nuclear charge, $Z_ie$ is the charge of the ion, $e$ is
proton charge. 
The formula (\ref{eq:EN}) can be obtained in a very simple way. The
second Newton law for the ion and its nucleus in the electric filed reads
\begin{eqnarray}
  M_i a_i = Z_i e E_0, \nonumber \\
  M_N a_N = Z e E_N, \nonumber
\label{eq:Mi}
\end{eqnarray}
where $M_i$ is the ion's mass,  $a_i$ is its acceleration,
$M_N$ is nuclear mass ($M_N \approx M_i$), and $a_N$ is its
acceleration. Since
the ion and its nucleus move together, the accelerations must be the
same ($a_i=a_N$),
therefore
\begin{equation}
 E_N = \frac{Z_i}{Z}E_0\frac{M_N}{M_i} \approx  \frac{Z_i}{Z} E_0. 
\label{eq:E_N}
\end{equation}
Different derivation of this formula can be found in Ref.~\cite{DFSS88,Oshima}.
Numerical calculations of the screened electric field inside an atomic
ion were performed in a number of our works (see, e.g. Ref.~\cite{DFSS88}). 

The Hamiltonian of the nuclear EDM ($d_N$) interaction with
the electric field is given by
\begin{equation}
  \hat H_d = d_N E_N = d_N\frac{Z_i}{Z}E_0.
\label{eq:Hd}
\end{equation}
Screening is stronger for diatomic molecules where we have an
additional suppression factor in  eq. (\ref{eq:E_N}),
$M_N/M_i=M_1/(M_1+M_2)$, 
where  $M_1$ and $M_2$ are the masses of the first and second
nucleus. 
For the average electric field acting on the first nucleus we obtain
\begin{equation}
 E_{1N}=\frac{Z_i}{Z_1}\frac{M_1}{M_1+M_2} E_0. 
\label{eq:e1n}
\end{equation}

Note that  the screeing factor here contains both nuclear masses. This
indicates that the nuclear motion can not be ignored and the screening
problem is more complicated than in atoms. For example, in \ a naive ionic
model of a neutral polar molecule $A^+ B^-$, both ions
$A^+$ and $ B^-$ should be located in the area of
zero (totally screened) electric field since their avarage acceleration is
zero. This could make $A^+$ and $B^-$ EDM
unobservable even if they are produced by a nuclear Schiff moment or
electron EDM. In a more realistic molecular calculations the Schiff moment
and electron EDM effects are not zero, however, they may be significantly
suppressed (in comparison with a naive estimate of ionic EDM
in  a very strong field of another ion) and  the results of the calculations
may be unstable.


In the case of monochromatic external electric field its frequency can
be chosen to be in resonance with the 
atomic electron excitation energy. 
Then for the effective Hamiltonian we can have
\begin{equation}
  \hat H_d = d_N E_N(t) \gg d_NE_0.
\label{eq:Hdt}
\end{equation}

\subsection{Electron EDM}

Paramagnetic atoms and molecules which have an unpaired electron are
most sensitive to the electron EDM. The EDM of such systems can be
expressed in the form
\begin{equation}
  d = K d_e,
\label{eq:d}
\end{equation}
where $d$ is the EDM of an atom or molecule, $d_e$ is electron EDM,
and $K$ is electron structure factor which comes from atomic
calculations. The factor $K$ increases with nuclear charge $Z$ 
as $Z^3$~\cite{Sandars65} times large relativistic factor
$R(Z\alpha)$~\cite{Flambaum76} which may exceed the value of 3 in heavy
atoms. A rough estimate of the enhancement factor 
in heavy atoms with external $s_{1/2}$ or $p_{1/2}$ electron is 
$K \sim 3Z^3\alpha^2R(Z\alpha) \sim 10^2 -
10^3$~\cite{Sandars65,Flambaum76}. 

Several orders of magnitude larger $K$ 
$\sim 10^7 -10^{11}$ exist in molecules due to the
mixing of the close rotational levels of opposite parity (including 
$\Lambda$-doublets)~\cite{SF78}. Following Sandars
this enhancement factor is usually presented as a ratio of a  very large
internal molecular field to the external electric field which polarizes the
molecule.

The best current limit on electron EDM comes from
the measurements of the thallium EDM~\cite{TlEDM} and reads
\begin{equation}
  d_e=(6.9 \pm 7.4) \times 10^{-28} e \ {\rm cm}.
\label{eq:deTl}
\end{equation}
Here the value $K=-585$~\cite{Liu} were used for the interpretation of
the measurements. The value of $K$ for Tl is very sensitive to the
inter-electron correlations but two most complete
calculations~\cite{Liu,DF09} give very close results. 

In contrast to paramagnetic atoms the diamagnetic (closed shell) atoms
are much less sensitive to electron EDM. This is because the only
possible direction of the atomic EDM in this case is along nuclear
spin and hyperfine structure interaction must be involved to link
electron EDM to nuclear spin. For example, for the mercury atom $K \sim
10^{-2}$~\cite{FK85,MP87}.  However, due to very strong constrain on
the mercury EDM~\cite{HgEDM} the limit on electron EDM extracted from
these measurements is competitive to the Tl result (\ref{eq:deTl})~\cite{HgEDM}
\begin{equation}
  d_e <  3 \times 10^{-27} e \ {\rm cm}.
\label{eq:deHg}
\end{equation}

New experiments are in progress to measure the electron EDM in
Cs~\cite{Gould}, Fr~\cite{FrEDM}, YbF~\cite{YbF}, ThO~\cite{ThO},
PbO~\cite{PbO}  and
in solid-state experiments~\cite{Sushkov}.

\subsection{Schiff moment}

Schiff moment is the lowest-order $T,P$-odd nuclear moment which appears
in the expansion of the nuclear potential when screening of the
external electric field by atomic electrons is taken into account.
This potential can be written as (see the derivation, e.g. in \cite{SAF,SAF08})
\begin{equation}
\phi (\mathbf{R}) = \int \frac{e\rho(r)}{|\mathbf{R}-\mathbf{r}|}d^3r
+ \frac{1}{Z}\left(\mathbf{d}\cdot\mathbf{\nabla}\right) 
\int \frac{\rho(r)}{|\mathbf{R}-\mathbf{r}|}d^3r,
\label{eq:phi}
\end{equation}
where $\rho(\mathbf{r})$ is nuclear charge density normalized to $Z$,
and $\mathbf{d}$ is 
nuclear EDM. The second term in (\ref{eq:phi}) is screening. 
Taking into account finite nuclear size the
lowest-order term in the expansion of (\ref{eq:phi}) in powers of $R$
can be written as~\cite{FG02}
\begin{equation}
\psi(\mathbf{R}) = - \frac{3\mathbf{S}\cdot\mathbf{R}}{B} \rho(R),
\label{eq:psiS}
\end{equation}
where $B=\int \rho(R)R^4dr$ and
\begin{equation}
  \mathbf{S} = \frac{e}{10}\left[ \langle r^2\mathbf{r}\rangle -\frac {5}{3Z}
    \langle r^2 \rangle \langle \mathbf{r} \rangle \right]
\label{eq:S}
\end{equation}
is Schiff moment. The expression (\ref{eq:psiS}) has no singularities
and can be used in relativistic calculations. The Schiff moment is
caused by the $T,P$-odd nuclear forces. The dominant mechanism is
believed to be $T,P$-odd nucleon-nucleon interaction. Another important
contribution comes from the EDMs of protons and neutrons. 

Schiff moment is the dominant nuclear contribution to the EDM of diamagnetic
atoms and molecules. The best limit on the EDM of diamagnetic atoms
comes from the measurements of the EDM of mercury performed in
Seattle~\cite{HgEDM} 
\begin{equation}
  |d(^{199}{\rm Hg})| < 3.1 \times 10^{-29} |e| {\rm cm}.
\label{eq:HgEDM}
\end{equation}
Interpretation of the measurements requires atomic and nuclear
calculations. Atomic calculations link the EDM of the atom to its
nuclear Schiff moment. Nuclear calculations relate Schiff moment to
the parameters of the $T,P$-odd nuclear interactions. Summary of
atomic~\cite{DFGK02,DFG07,DFP09} and nuclear~\cite{FKS86,octSM}
calculations for diamagnetic atoms of experimental 
interest is presented in Table~\ref{t:EDM}. 
To compare the EDM of different atoms we present only the results of our
nuclear calculations which all were performed by the same
method. For Hg and Ra there are several recent nuclear many-body
calculations available (see references in the most recent calculation
\cite{Ban}) and new calculations are in progress.

The dimensionless constant
$\eta$ characterizes the strength of the $P,T$-odd nucleon-nucleon
interaction which is to be determined from the EDM measurements. Using
(\ref{eq:HgEDM}) and the data from the Table one can get
\begin{equation}
  S(^{199}{\rm Hg}) = (-1.8 \pm 4.6 \pm 2.7) \times 10^{-13} e \ {\rm
    cm}
\label{eq:SHg}
\end{equation}
and for the $T,P$-odd neutron-proton interaction
\begin{equation}
  \eta_{np} = (1\pm 3 \pm 2) \times 10^{-5}.
\label{eq:eta}
\end{equation}

\begin{table}
\caption{EDMs of diamagnetic atoms of experimental interest.}
\label{t:EDM}
\begin{ruledtabular}
\begin{tabular}{rr ccl}
\multicolumn{1}{c}{$Z$} & 
\multicolumn{1}{c}{Atom} & 
\multicolumn{1}{c}{$[S/(\ e \ {\rm fm}^3)]$} &
\multicolumn{1}{c}{$\eta \ e \ {\rm cm}$} &
\multicolumn{1}{c}{Experiment}  \\
&&\multicolumn{1}{c}{$\times 10^{-17}e \ {\rm cm}$} &
\multicolumn{1}{c}{$\times 10^{-25}$} & \\

\hline
 2 & $^3$He    & $8\times 10^{-5}$ & $5\times10^{-4}$ & \\
54 & $^{129}$Xe & 0.38 & 0.7 & Seattle~\cite{Seattle}, Ann
Arbor~\cite{AnnArbor} \\
&&&& Princeton~\cite{Princeton} \\
70 & $^{171}$Yb & -1.9 & 3 & Bangalore~\cite{Bangalore},
Kyoto~\cite{Kyoto} \\
80 & $^{199}$Hg & -2.8 & 4 & Seattle~\cite{HgEDM} \\
86 & $^{223}$Rn & 3.3 & 3300 & TRIUMF~\cite{RnEDM} \\
88 & $^{225}$Ra & -8.2 & 2500 & Argonne~\cite{Argonne}, KVI~\cite{KVI} \\
88 & $^{223}$Ra & -8.2 & 3400 &  \\
\end{tabular}
\end{ruledtabular}
\end{table}

\paragraph{Nuclear  enhancement.} 
It was pointed out in Ref.~\cite{octSM}
that Schiff moment of nuclei
with octupole deformation can be strongly enhanced. This can be
explained in a very simple way. Nuclear deformation creates an
intrinsic Schiff moment in the nuclear reference frame
\begin{equation}
  S_{intr} \approx eZR_N^3 \frac{9\beta_2\beta_3}{20\pi\sqrt{35}},
\label{eq:Sintr}
\end{equation}
where $R_N$ is nuclear radius, $\beta_2 \approx 0.2$ is the parameter
of quadrupole deformation, and $\beta_3 \approx 0.1$ is the parameter
of octupole deformation. The intrinsic Schiff moment (\ref{eq:Sintr})
does not violate $T$ or $P$ invariance and, if no $T,P$-odd
interaction present, it averages to zero in the
laboratory reference frame due to nuclear rotation. However, when
$T,P$-odd interaction is included, it can mix close rotational states
of opposite parity. Small energy interval between these states leads
to strong enhancement of the nuclear Schiff moment in the laboratory
reference frame
\begin{eqnarray}
\label{eq:Sbig}
 && S_{lab} \sim \frac{\langle + | H_{PT} | - \rangle}{E_+ - E_-}
  S_{intr} \sim \\
&&0.05e\beta_2\beta_3^2ZA^{2/3}\eta r_0^3\frac{\rm eV}{E_+ - E_-} \sim
700\times 10^{-8} \eta e \ {\rm fm^3}, \nonumber
\end{eqnarray}
where $r_0=1.2$ fm is the inter-nucleon distance $|E_+ - E_-| \sim$ 50
keV. The estimate (\ref{eq:Sbig}) is about 500 times larger than the
Schiff moment of a spherical nucleus like Hg.

It was pointed out in Ref.~\cite{softSM} that octupole deformation
doesn't need to be static. Soft octupole vibrations lead to similar
enhancement. Large values of the Schiff moment for Ra and Rn (see
Table~\ref{t:EDM}) are due to nuclear octupole deformation.

\paragraph*{Acknowledgments.} 
This work was supported in part by the Australian Research Council and ECT.

\end{document}